\documentclass[runningheads]{llncs}
\usepackage[T1]{fontenc}
\usepackage{amsmath}
\usepackage{newtxtext}
\usepackage[varvw]{newtxmath}
\usepackage{graphicx,booktabs}
\usepackage{tikz}
\usetikzlibrary{arrows.meta}
\usepackage[hidelinks]{hyperref}
\usepackage[misc]{ifsym}

\newcommand{\cat}{\mathcal{I}}
\newcommand{\valid}{\mathcal{C}}
\newcommand{\vocab}{\mathcal{V}}
\newcommand{\hist}{H_u}
\title{Jev-GR: Autoregressive Semantic ID Prediction with a Decision LLM for Generative Recommendation}
\titlerunning{Jev-GR for Semantic ID Prediction}
\author{Jingyu Li\inst{1}\thanks{Jingyu Li and Chenhao He contributed equally to this work.} \and Chenhao He\inst{2\star} \and
Qiuyi Xin\inst{2} \and Pengcheng Zhang\inst{3,4}}
\authorrunning{J. Li et al.}
\institute{College of Computer Science and Technology, Tianjin University, Tianjin 300354, China\\
\email{lijingyu\_working@tju.edu.cn}
\and College of Computer Software, Tianjin University, Tianjin 300354, China\\
\email{\{chenhao4617, qiuyi\}@tju.edu.cn}
\and Key Laboratory of Water Big Data Technology of Ministry of Water Resources, Nanjing 211100, China
\and The College of Computer Science and Software Engineering, Hohai University, Nanjing 211100, China\\
\email{pchzhang@hhu.edu.cn}}
\begin{document}
\maketitle

\begin{abstract}
Large language models (LLMs) have spurred interest in generative recommenders that use item descriptions alongside interaction histories to predict what a user will choose next. Generative recommenders retrieve items by predicting semantic identifiers (SIDs) one symbol at a time. Users who reach the same SID prefix face the same catalog-valid continuations, yet their histories can favor different subtrees. An observed item's path supplies a ranking signal only at prefixes with competing continuations. We propose Jev-GR, which encodes a user's historical SIDs, their corresponding item descriptions, and the current prefix with a Qwen backbone and an OpenJev-initialized adapter. A depth-specific projection scores the valid next symbols from one shared prefix state. Supervised fine-tuning averages each target's loss over its branching positions, and beam search ranks unseen items by the cumulative log probabilities of their paths. On the Amazon Industrial and Scientific and Tools and Home Improvement (5-core) datasets, Jev-GR achieves Recall@20 of 0.0893 and 0.0453, respectively, compared with 0.0721 and 0.0349 for the strongest baseline on each dataset. Ablation study shows that Recall and NDCG fall on both datasets when the adapter is frozen or the loss is normalized by the full SID length.

\end{abstract}

\section{Introduction}
Recent works on large language models (LLMs) for recommendation have explored how the text associated with a user's interactions, such as product titles and descriptions, can reveal preferences that item IDs alone leave implicit~\cite{wu2024llmrec}. In parallel, generative recommendation has gained attention as a way to retrieve items by generating their identifiers from a user's history~\cite{li2024generativesearch}. The two approaches meet when a language model reads a history of described items and predicts a code for the next one. If related items share the first symbols of their codes, each prediction narrows the set of items still in contention.

\begin{figure}[htbp]
\centering
\includegraphics[width=0.75\linewidth]{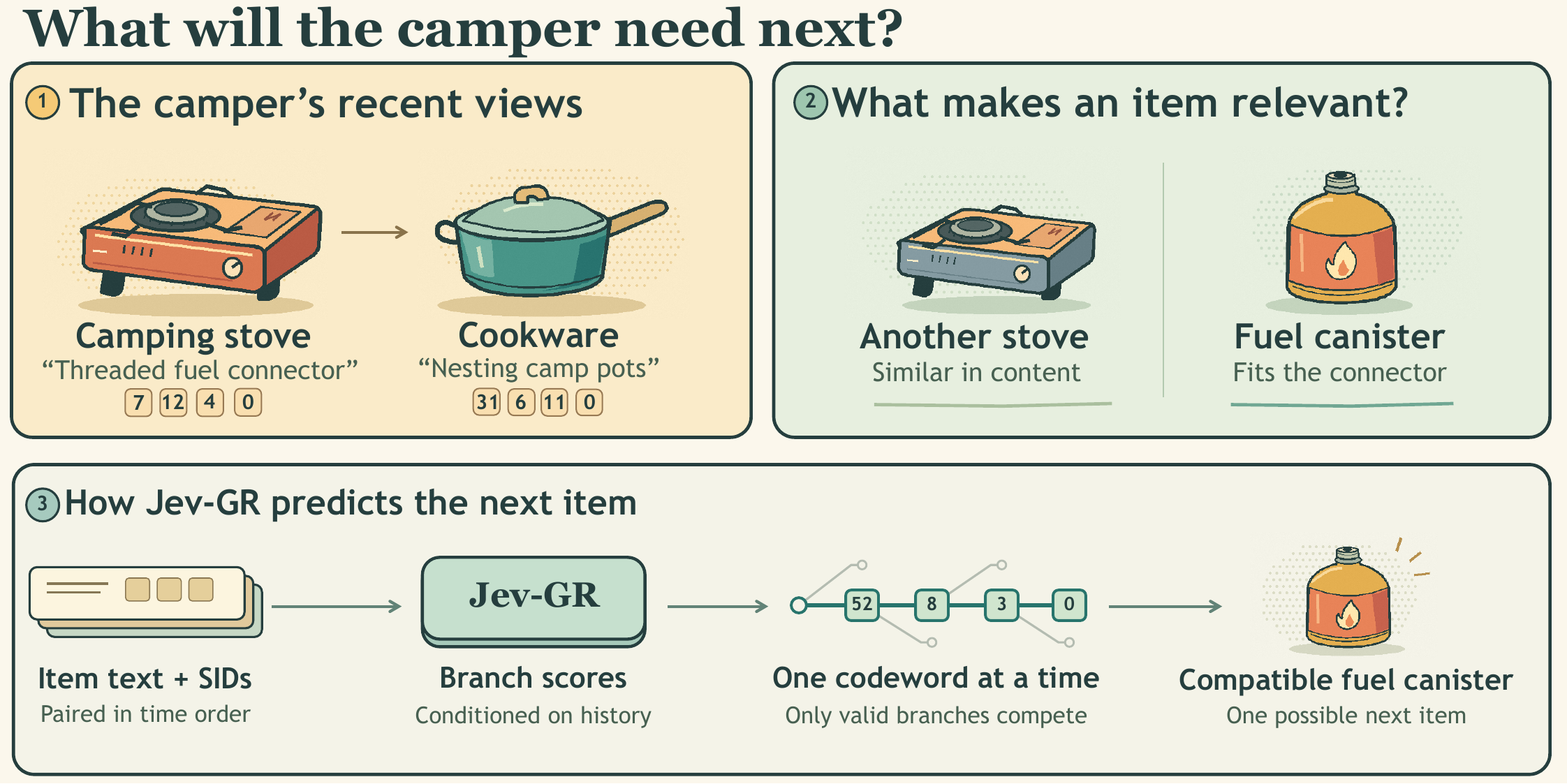}
\caption{Illustrative shopping history and semantic identifiers (SIDs).}
\label{fig:introduction}
\end{figure}

A semantic identifier (SID) gives the recommender a path to an item~\cite{rajput2023tiger}. Items can share a prefix because their descriptions are similar, even when the next item a user needs serves a complementary purpose~\cite{tsaruk2026preference}. Consider the shopper in Fig.~\ref{fig:introduction} who has viewed a camping stove and cookware. Another stove may be close to the first in the catalog's content index, while a fuel canister that fits its threaded connector would let the shopper use the stove. The historical SIDs locate the products the shopper viewed, and the stove description names the connector. The recommender must learn when that detail favors a branch containing compatible fuel.

At a given prefix, every user faces the same legal continuations. A Transformer trained on SID sequences can predict these continuations from interaction histories~\cite{rajput2023tiger}, and a general-purpose LLM can be fine-tuned to generate catalog-valid paths~\cite{kong2025minionerec}. Aligning item indices with language or supplementing them with metadata and continuous representations can give the model more evidence about the user~\cite{zheng2024lcrec,lee2025gram,liu2026scaling}. Our first challenge is to turn that evidence into a ranking of the legal branches at this fork. Decision LLMs such as Jev assign probabilities to options supplied with a context~\cite{almeida2026jev,typesafechoice}. OpenJev is trained to read each candidate with the context before assigning it a score~\cite{openjev,openjevcard}. If this training helps identify which details matter to a choice, some of that evidence may be recoverable from a single representation of the history and prefix, shared by all siblings.

A second challenge is to learn branch preferences from an interaction that identifies only the final product. At a prefix with one legal continuation, every shopper who reaches it must take the same step. At a fork, the observed interaction identifies one subtree among several that were still reachable. Two product SIDs of equal length can therefore contain different numbers of decisions that provide evidence about preference.

We introduce \emph{Jev-GR} to learn those comparisons. Residual $k$-means (RQ-KMeans) encodes item-text embeddings into successive codewords~\cite{ju2025handbook}. A Qwen backbone with an adapter initialized from OpenJev reads the historical SIDs, their descriptions in the same order, and the current prefix. At each fork, a depth-specific projection uses this shared state to assign probabilities to the legal children. During supervised fine-tuning (SFT), we update the LoRA adapter~\cite{hu2022lora}, SID embeddings, and projections, averaging each target's loss over its branching positions. At retrieval time, beam search accumulates the log probabilities of successive choices to rank unseen items.

Our contributions are threefold.
\begin{itemize}
\item We formulate next-SID prediction as a test of whether per-candidate decision training transfers to personalized sibling rankings. We specify a comparison of the released and freshly initialized adapters with the Qwen backbone, scoring rule, and catalog held fixed.
\item We develop a recommender that uses paired text and SID histories to score an item path at its forks. To isolate the contribution of that pairing, we propose comparing SID-only, text-only, and paired histories with the same target codes. Permuting complete codes across items would separately test the value of the catalog grouping.
\item On Amazon Industrial and Scientific and 2014 Tools and Home Improvement (5-core) dataset~\cite{ni2019amazon}, Jev-GR achieves the best through several vital recommendation baselines on 6 metrics and 2 dataset.
\end{itemize}

\section{Related Work}

\subsection{Language models for decisions.}
TypeSafe introduced Jev as a System One model for mapping unstructured input to typed decisions. Its release describes reinforcement learning for calibrated decisions as the training approach~\cite{almeida2026jev}. The \emph{Choice} primitive exposes a probability for each caller-supplied option~\cite{typesafechoice}. TypeSafe's hierarchical classification example already composes such choices down a taxonomy, using beam search to retain several candidate leaves~\cite{typesafehierarchy}. It classifies a document among named categories. In sequential recommendation, two users at the same node can prefer different items, and the children are learned quantization symbols. Their interpretation must be acquired from interactions. JevBench independently evaluates typed decision accuracy, calibration, latency, and cost~\cite{jevbench}. Here the outcome of interest is the rank of the user's held-out item after several dependent choices.

\subsection{Open decision backbones.}
OpenJev implements candidate scoring with a language backbone, a low-rank adapter, and a scalar head. The released 2B implementation fits candidate distributions with cross-entropy and a Brier term~\cite{openjev,openjevcard}. Jeva~\cite{li2026jeva} is a native System One model that independently encodes requests and candidates. Its architecture supports null awareness, permutation-equivariant candidate probabilities, variable-size candidate sets, and cross-request reuse of candidate representations. Jev-GR transfers its adapter and learns a new readout for each SID depth. A matched Qwen initialization then permits a direct test of what the earlier decision training contributes. This comparison concerns the released OpenJev weights and the recommendation objective used here.

\subsection{Generative retrieval and language-based ranking.}
TIGER establishes next-item prediction through autoregressive SID generation~\cite{rajput2023tiger}. LETTER asks how semantic, collaborative, and diversity objectives shape the item codes~\cite{wang2024letter}. MiniOneRec studies Qwen adaptation and catalog-constrained generation~\cite{kong2025minionerec}. Hou et al.\ explicitly frame recommendation as a language-model ranking problem and rank a list of complete items supplied by candidate generators~\cite{hou2024rank}. These studies address different parts of our question. A constrained generator already produces legal paths, and a language ranker already compares alternatives. We study decision training for a different ranking unit. Each alternative is the next SID symbol, and selecting it determines which items remain reachable.

\subsection{Ranking along a tree.}
Hierarchical probabilistic language models factor a leaf probability into local decisions along a path~\cite{morin2005hierarchical}. Tree-based recommendation uses an item hierarchy to organize retrieval from a large catalog~\cite{zhu2018tdm}. Of particular relevance to our scorer, the Neural Corpus Indexer generates document IDs using a prefix-aware, weight-adaptive decoder~\cite{wang2022nci}. In our model all prefixes at a given depth share one projection matrix. The prefix affects the backbone state and determines which rows of that matrix compete in the softmax.

\subsection{Where a path becomes difficult.}
DASO locates levels at which sampled continuations diverge from a target SID and uses the observed prefix-match depth to guide training~\cite{yu2026daso}. That difficulty can change as the policy learns. The branching count used here is a property of a given catalog and target path. A 20-way branch can be easy for a particular user, while a two-way branch can separate two equally plausible items. The catalog count determines whether a training position supplies a ranking gradient. The model's errors then show which of those decisions it has learned.

\section{Preliminaries: Catalog and Prediction Task}
\label{sec:catalog}

Let $\cat$ be the item catalog, $x_i$ the static text of item $i$, and $\hist=(i_1,\ldots,i_n)$ the ordered history available before user $u$'s next interaction. Write $H_u^{(L)}$ for the suffix of length $\min(L,n)$ and $\mathcal{O}_u=\{i_j:1\le j\le n\}$ for the set of previously observed items. The recommender conditions on $H_u^{(L)}$ and ranks items in $\cat\setminus\mathcal{O}_u$. Each training pair $(\hist,i^\star)$ supplies an observed next item $i^\star$.

\subsection{Constructing semantic identifiers}

Given a history $\hist$ and its observed next item $i^\star$, semantic-ID recommendation predicts a target sequence $s^\star=s(i^\star)$ from a fixed catalog index~\cite{rajput2023tiger,kong2025minionerec,yu2026daso}. To construct this index, we first encode the description $x_i$ of each item using residual $k$-means (RQ-KMeans). Its semantic code $q(i)=(s_1(i),\ldots,s_Q(i))$ contains one centroid index $s_t(i)\in\{0,\ldots,J-1\}$ per depth. When several items share a code, an identity suffix lets the recommender distinguish them.

\paragraph{Learning the semantic code.}
A frozen text encoder produces one vector $v_i$ per item by mean pooling the hidden states of its non-padding tokens and normalizing the result to unit length. Set $r_{i,0}=v_i$. At depth $t$, the input to $k$-means is the residual $r_{i,t-1}$ left by the preceding codebooks. With those codebooks fixed, we fit $E_t=(e_{t,0},\ldots,e_{t,J-1})$ to the catalog by minimizing
\begin{equation}
 \mathcal{L}_{\mathrm{RQ}}^{(t)}(E_t)
 = \frac{1}{|\cat|}\sum_{i\in\cat}
       \min_{0\le j<J}\bigl\|r_{i,t-1}-e_{t,j}\bigr\|_2^2 .
 \label{eq:rq-loss}
\end{equation}
Each iteration assigns residuals to their nearest centroids, then updates each occupied centroid to the mean of its assigned residuals. Once the codebook is fitted, the centroid index gives the next semantic symbol and the remaining error passes to the next depth,
\begin{equation}
 s_t(i)=\arg\min_{0\le j<J}\bigl\|r_{i,t-1}-e_{t,j}\bigr\|_2^2,\qquad
 r_{i,t}=r_{i,t-1}-e_{t,s_t(i)},\qquad 1\le t\le Q .
 \label{eq:rq}
\end{equation}
Items with the same first $t$ symbols thus share the partial reconstruction $\sum_{\ell=1}^{t}e_{\ell,s_\ell(i)}$. A later symbol refines that approximation. The recommender gives each pair $(t,s_t(i))$ its own input token because index $j$ refers to a different codebook at each depth.

\paragraph{From codes to prediction targets.}
Let $G(q)=\{i\in\cat:q(i)=q\}$. When any group contains multiple items, we append one symbol to every code, using the item's zero-based rank within $G(q)$ in item-ID order. Otherwise the semantic code itself identifies the item. The resulting paths have a common depth $D=Q+\delta$, where $\delta$ is one if the index uses a suffix and zero otherwise. We fix this item-to-SID mapping before fitting the recommender. For $s^\star_{<t}=(s^\star_1,\ldots,s^\star_{t-1})$, the model assigns the target item the path probability
\begin{equation}
 P_\theta\!\left(s^\star\mid\hist\right)
   =\prod_{t=1}^{D}P_\theta\!\left(s^\star_t\mid\hist,s^\star_{<t}\right).
 \label{eq:sid-path}
\end{equation}
At the root the prefix is empty. After each choice, only items that share the extended prefix remain reachable. The catalog therefore determines which symbols can compete in each factor of Eq.~\eqref{eq:sid-path}.

\subsection{Catalog branches}

Write $p\preceq s$ when $p$ is a prefix of $s$, and define its descendant items as $\cat(p)=\{i\in\cat:p\preceq s(i)\}$. For a nonterminal prefix $p$ at depth $t-1$, the available actions are the symbols
\begin{equation}
 \valid(p)=\{s_t(i):i\in\cat(p)\}.
 \label{eq:support}
\end{equation}
The sets $\cat(p||c)$ for $c\in\valid(p)$ partition $\cat(p)$, where $p||c$ denotes appending concatenation from $c$ to $p$. Choosing a symbol therefore selects the subtree in which retrieval continues. A prefix with a single child passes its entire probability mass to that child.

\section{Methodology: Autoregressive SID Prediction with Jev-GR}
\label{sec:method}

\begin{figure}[tb]
\centering
\includegraphics[width=1.0\linewidth]{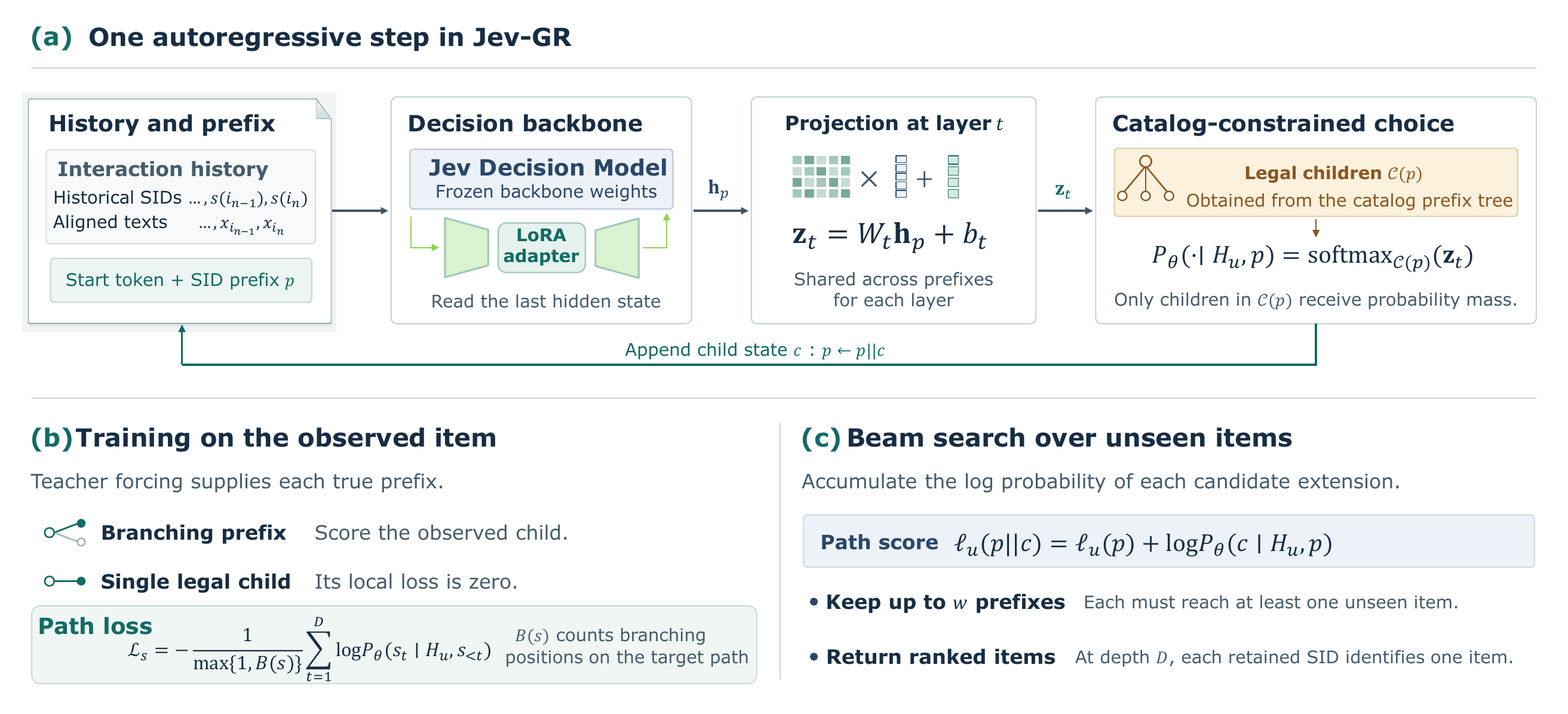}
\caption{Autoregressive SID prediction in Jev-GR. A local choice scores the legal children of a prefix. Each retained child becomes part of the state that determines the next choice.}
\label{fig:decoder}
\end{figure}

For two users at the same prefix $p$, the catalog supplies the same children $\valid(p)$, but the users' histories can favor different subtrees. Jev-GR models each factor in Eq.~\eqref{eq:sid-path} as a choice among these children. Its backbone encodes the user's history and the prefix once, and its output projection compares the legal continuations from that shared state (Fig.~\ref{fig:decoder}). An observed next item identifies one child at every prefix on its path. The number of those prefixes that offer a competing child determines how the path contributes to training.

\subsection{Transferring OpenJev to a shared prefix}

In its source task, OpenJev appends each candidate to the context and reads a scalar score from the resulting candidate-specific state~\cite{openjevcard}. Jev-GR retains the Qwen backbone and initializes its low-rank adapter $\Delta$ from that decision checkpoint. It moves the comparison among candidates to a depth-specific output projection. The prefix state must therefore encode the evidence for ranking all children before any of their symbols enters the backbone.

\begin{figure}[htbp]
\centering
\includegraphics[width=0.75\linewidth]{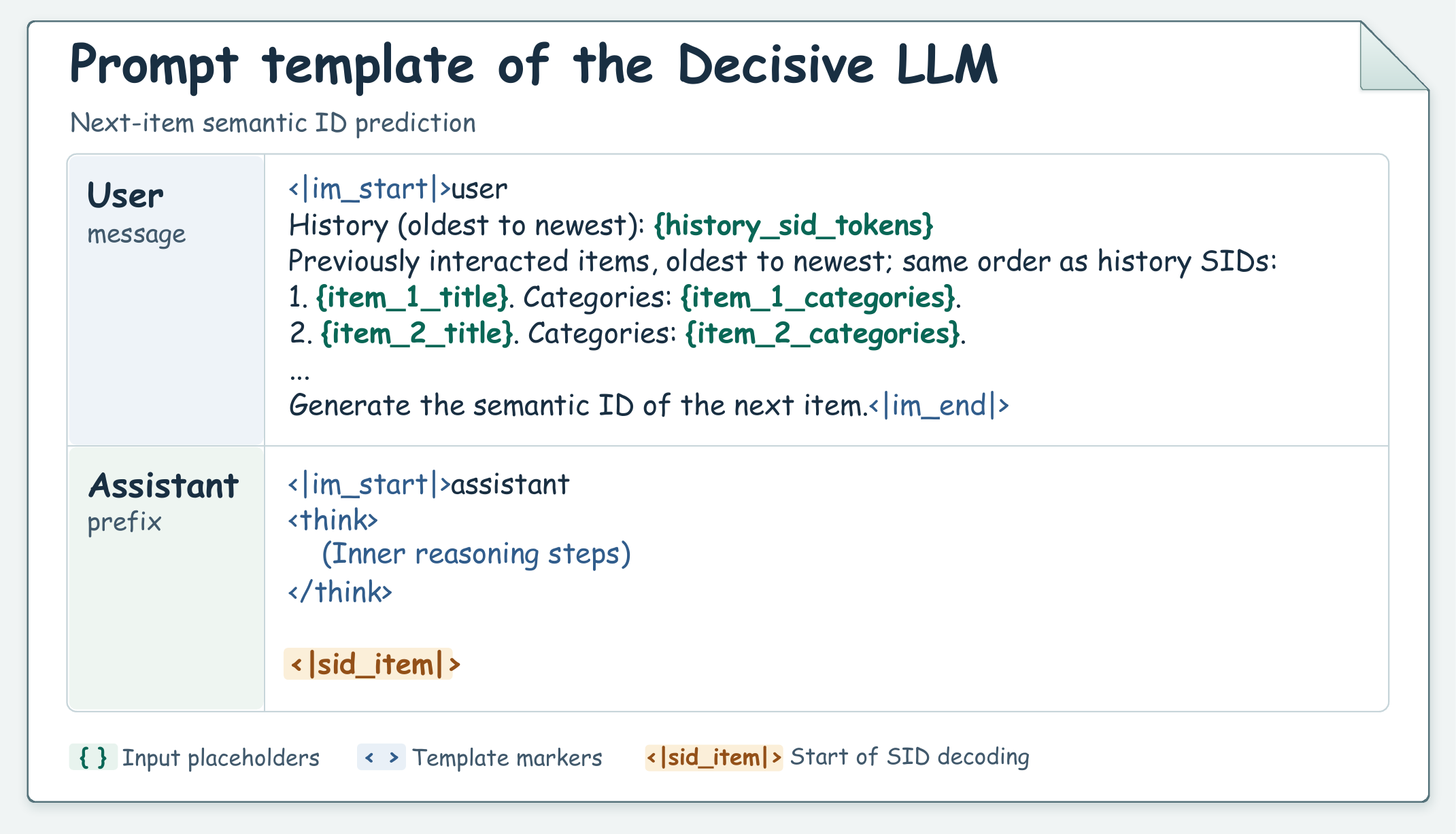}
\caption{Prompt template of the Decisive LLM (OpenJev) model.}
\label{fig:prompt}
\end{figure}

For user $u$, we form a prompt $\pi_u$ from the recent history $H_u^{(L)}$ (Fig.~\ref{fig:prompt}). It lists the historical SIDs in chronological order with item boundaries, followed by the descriptions of those same items in the same order. The SIDs identify the catalog paths the user has followed. The descriptions can identify attributes that matter across purchases, such as the connector required by a previously viewed stove. Write $\tau(p)$ for the sequence of depth-specific input tokens representing $p$. For fixed $u$, the state used to score its children is
\begin{equation}
 h_p=\operatorname{last}\!\left(
 f_{\Theta_0,\Delta}\bigl(\pi_u\Vert\tau_{\mathrm{start}}\Vert\tau(p)\bigr)
 \right)\in\mathbb{R}^{d}.
 \label{eq:state}
\end{equation}
Here $\Vert$ denotes concatenation and $\operatorname{last}$ selects the hidden state at the final input position. At the empty prefix, $h_p$ is the state of the start-of-item token $\tau_{\mathrm{start}}$. At later depths it is the state of the most recently appended SID token.

\subsection{Ranking the children of a prefix}

At a prefix of length $t-1$, each possible next symbol $c$ has a row $w_{t,c}^{\top}$ and bias $b_{t,c}$ in the projection for depth $t$. Let $\vocab_t=\{s_t(i):i\in\cat\}$ denote the symbols present at that depth. The projection produces logits $z_t=W_th_p+b_t$ over $\vocab_t$. The catalog restricts the normalization to $\valid(p)$ from Eq.~\eqref{eq:support},
\begin{equation}
 P_\theta(c\mid\hist,p)=
 \begin{cases}
 \displaystyle\frac{\exp(z_{t,c})}{\sum_{j\in\valid(p)}\exp(z_{t,j})},
    & c\in\valid(p),\\[4pt]
 0, & c\notin\valid(p).
 \end{cases}
 \label{eq:local}
\end{equation}
Conditioning on $\hist$ in this distribution uses the history window in $\pi_u$. All prefixes at depth $t-1$ share $W_t$ and $b_t$. For siblings $a,b\in\valid(p)$, the common denominator cancels, giving
\begin{equation}
 \log\frac{P_\theta(a\mid\hist,p)}{P_\theta(b\mid\hist,p)}
 =(w_{t,a}-w_{t,b})^\top h_p+b_{t,a}-b_{t,b}.
 \label{eq:odds}
\end{equation}
Thus, at a given fork, two users can rank the same siblings in opposite orders if their states give opposite signs in Eq.~\eqref{eq:odds}. Where the source scorer forms a suffix representation for each candidate, this projection uses one backbone state for all $|\valid(p)|$ siblings. A selected symbol enters the backbone when it becomes part of the prefix for the next choice.

\subsection{Supervised fine-tuning at catalog forks}

An observed item supplies the target path $s=s(i^\star)$. At depth $t$, teacher forcing conditions on the true prefix $s_{<t}$ and assigns the local loss $-\log P_\theta(s_t\mid\hist,s_{<t})$. If the target symbol is the only legal child, Eq.~\eqref{eq:local} gives it probability one for every setting of the model parameters. Its local loss is zero, although the symbol still conditions later choices. The number of positions on the path that can contribute a ranking gradient is:
\begin{equation}
 B(s)=\sum_{t=1}^{D}\mathbf{1}\!\left\{|\valid(s_{<t})|>1\right\}.
 \label{eq:branch}
\end{equation}
For a minibatch $\mathcal{B}$ of histories and target paths $(H,s)$, Jev-GR minimizes:
\begin{equation}
 \mathcal{L}(\theta)=-\frac{1}{|\mathcal{B}|}
 \sum_{(H,s)\in\mathcal{B}}
 \frac{\sum_{t=1}^{D}\log P_\theta(s_t\mid H,s_{<t})}{\max\{1,B(s)\}} .
 \label{eq:loss}
\end{equation}
Each example with $B(s)>0$ contributes its mean loss per branching decision. Without this normalization, a path with more forks would contribute more terms even at the same mean prediction error. A path with no fork contributes zero.

Appending $s_{<D}$ to the prompt after the start token supplies all teacher-forced states in one causal pass. We freeze $\Theta_0$ and train $\theta=\{\Delta,E_{\mathrm{SID}},(W_t,b_t)_{t=1}^{D}\}$, where $E_{\mathrm{SID}}$ contains the trainable SID input embeddings. The LoRA adapter uses rank $\rho$ and scaling parameter $\alpha$~\cite{hu2022lora}. AdamW updates the adapter, embeddings, and projections with respective learning rates $\eta_\Delta,\eta_E,\eta_W$, weight decay $\lambda$, and gradient-norm clipping threshold $\gamma$.

\subsection{Retrieving unseen items by path probability}

At inference each retained prefix supplies its own context for the next decision. To rank complete items by Eq.~\eqref{eq:sid-path}, we start at the empty prefix with log score $\ell_u(\epsilon)=0$ and accumulate
\begin{equation}
 \ell_u(p\Vert c)=\ell_u(p)+\log P_\theta(c\mid\hist,p),\qquad
 S_\theta(i\mid\hist)=\ell_u\bigl(s(i)\bigr).
 \label{eq:rank}
\end{equation}
An extension is eligible if its subtree contains an unseen item, that is, if $\cat(p\Vert c)\setminus\mathcal{O}_u\ne\varnothing$. We first compute each local probability over all catalog children, then filter fully seen subtrees, preserving the original scores of the eligible extensions. Across all parents in the current beam, we retain up to $w$ eligible extensions with the largest cumulative scores. At depth $D$, $S_\theta(i\mid\hist)=\log P_\theta(s(i)\mid\hist)$ for each surviving item. The index maps each leaf to its unseen item, and sorting these items by score gives the recommendation list.

\section{Experiments}
\label{sec:experiments}

We examine how well Jev-GR retrieves unseen items, which gains depend on adapting OpenJev and weighting catalog forks, and how retrieval changes with SID width and depth.

\subsection{Experimental setup}
\label{sec:settings}

\paragraph{Datasets and evaluation.}
We evaluate product retrieval on Amazon Industrial and Scientific~\cite{ni2019amazon} and the Tools and Home Improvement 5-core subset of Amazon Reviews 2014.\footnote{\url{https://jmcauley.ucsd.edu/data/amazon/}} In the released Tools 5-core subset, each user and item has at least five reviews. In both categories, we order interactions by time, retain the first occurrence of each user--item pair, and keep users with at least five distinct items. Each user contributes one training target, the fourth item from the end. The next three items are the calibration, validation, and test targets, respectively. Only earlier interactions are available at each prediction. The prepared Industrial cohort contains 9,331 users and 5,285 items.

All methods within a dataset use the same split and rank unseen items against the full catalog. We select checkpoints by validation NDCG@10. With one held-out target per user, Recall@$K$ is the fraction of users whose target appears in the first $K$ results. NDCG@$K$ assigns a hit at rank $r$ the weight $1/\log_2(1+r)$. We report both metrics at $K=10$ and $20$. Hits gives the corresponding number of successful users.

\paragraph{Baselines and implementation.}
SASRec~\cite{kang2018sasrec} and BERT4Rec~\cite{sun2019bert4rec} rank items from sequential histories. TIGER~\cite{rajput2023tiger} retrieves items by generating SIDs, and LETTER--TIGER tests a learned tokenization~\cite{wang2024letter} with the same baseline architecture. Jev-GR encodes item text with frozen Qwen3.5-2B and fits $Q=3$ residual codebooks of $J=128$ centroids each. Its frozen Qwen backbone has hidden width $d=2{,}048$. We initialize the adapter from OpenJev-2B~\cite{openjevcard,qwen} with LoRA rank $\rho=8$ and scale $\alpha=16$. The history window is $L=20$. AdamW uses learning rates $\eta_\Delta=2\times10^{-5}$ and $\eta_E=\eta_W=10^{-4}$, weight decay $\lambda=0.01$, and gradient-norm threshold $\gamma=1$. Validation and test beam widths are 10 and 20, respectively.

\subsection{Overall Performance}
\begin{table}[tb]
\caption{Full-catalog next-item retrieval. Scores are fractions. Bold marks the largest measured score within each complete dataset panel.}
\label{tab:industrial}
\centering\small
\begin{tabular}{@{}lrrrrrr@{}}
\toprule
 & \multicolumn{3}{c}{Top 10} & \multicolumn{3}{c}{Top 20}\\
\cmidrule(lr){2-4}\cmidrule(l){5-7}
Method & Hits & Recall & NDCG & Hits & Recall & NDCG\\
\midrule
\multicolumn{7}{l}{\emph{Industrial and Scientific:}}\\
\midrule
SASRec       & 554 & 0.0594 & 0.0439 & 673 & 0.0721 & 0.0472\\
BERT4Rec     & 203 & 0.0218 & 0.0140 & 308 & 0.0330 & 0.0169\\
TIGER        & 480 & 0.0514 & 0.0326 & 659 & 0.0706 & 0.0374\\
LETTER--TIGER& 452 & 0.0484 & 0.0317 & 597 & 0.0640 & 0.0357\\
Jev-GR       & \textbf{605} & \textbf{0.0648} & \textbf{0.0443} & \textbf{834} & \textbf{0.0893} & \textbf{0.0505}\\
\midrule
\multicolumn{7}{l}{\emph{Tools and Home Improvement (5-core):}}\\
\midrule
SASRec        & 367 & 0.0221 & 0.0127 & 518 & 0.0311 & 0.0150\\
BERT4Rec      & 321 & 0.0193 & 0.0111 & 539 & 0.0324 & 0.0144\\
TIGER         & 393 & 0.0236 & 0.0125 & 580 & 0.0349 & 0.0153\\
LETTER--TIGER & 362 & 0.0218 & 0.0112 & 532 & 0.0320 & 0.0138\\
Jev-GR        & \textbf{513} & \textbf{0.0308} & \textbf{0.0160} & \textbf{754} & \textbf{0.0453} & \textbf{0.0194}\\
\bottomrule
\end{tabular}
\end{table}

Table~\ref{tab:industrial} reports full-catalog next-item retrieval for Jev-GR ($J=128$, $Q=3$) and four baselines. Jev-GR ranks first in all eight Recall and NDCG comparisons across the two datasets and two cutoffs. On Industrial, it retrieves 605 targets in the top 10 and 834 in the top 20, exceeding SASRec by 51 and 161 hits, respectively. It also achieves the highest NDCG@10 (0.0443 versus 0.0439 for SASRec). On Tools, Jev-GR retrieves 513 and 754 targets at the same cutoffs, exceeding TIGER by 120 and 174 hits. The relative gains in Recall@20 over the strongest baseline are 23.9\% on Industrial (0.0893 versus 0.0721) and 29.8\% on Tools (0.0453 versus 0.0349).

Jev-GR scores all legal next symbols from one user-conditioned state per prefix and outperforms both SID-generation baselines on item-level Recall and NDCG at both cutoffs. Compared with TIGER, the stronger of these baselines, Jev-GR raises NDCG@20 from 0.0374 to 0.0505 on Industrial and from 0.0153 to 0.0194 on Tools. The training ablations in Table~\ref{tab:variants} show that fine-tuning the OpenJev-initialized adapter and averaging the loss over branching positions each improve Recall and NDCG on both datasets.

\subsection{Ablation study}

Table~\ref{tab:variants} compares two training ablations with full Jev-GR at $J=128$ and $Q=3$.

\subsubsection{Frozen adapter.}
We keep the OpenJev-initialized adapter fixed during supervised fine-tuning and train the SID embeddings and depth-specific projections. Freezing the adapter lowers Recall and NDCG at both cutoffs on both datasets. NDCG@20 falls from 0.0505 to 0.0290 on Industrial and from 0.0194 to 0.0135 on Tools, showing the benefit of adapting the OpenJev weights to next-item retrieval.

\subsubsection{Fixed-depth loss.}
We replace the normalization factor $\max\{1,B(s)\}$ in Eq.~\eqref{eq:loss} with the full SID length $D$. Positions with a single legal continuation contribute zero local loss, so dividing by $D$ gives less weight to paths with fewer forks. On Industrial, Recall@20 falls from 0.0893 to 0.0866 and NDCG@20 from 0.0505 to 0.0347. On Tools, Recall@20 falls from 0.0453 to 0.0435 and NDCG@20 from 0.0194 to 0.0170. The larger relative drops in NDCG indicate that normalization by the number of forks benefits discounted ranking more than top-20 coverage.

\begin{table}[tb]
\caption{Training ablations and SID width and depth comparisons on the two Amazon datasets. R and N denote Recall and NDCG. The first row is the reference for both sets of comparisons.}
\label{tab:variants}
\centering\small\setlength{\tabcolsep}{2.6pt}
\begin{tabular}{@{}lrrrrrrrr@{}}
\toprule
 & \multicolumn{4}{c}{Industrial} & \multicolumn{4}{c}{Tools 5-core}\\
\cmidrule(lr){2-5}\cmidrule(l){6-9}
Variant & R@10 & N@10 & R@20 & N@20 & R@10 & N@10 & R@20 & N@20\\
\midrule
Full ($J{=}128$, $Q{=}3$) & 0.0648 & 0.0443 & 0.0893 & 0.0505 & 0.0308 & 0.0160 & 0.0453 & 0.0194\\
\midrule
\multicolumn{9}{l}{\emph{Training ablations:}}\\
\midrule
Frozen adapter & 0.0537 & 0.0246 & 0.0715 & 0.0290 & 0.0234 & 0.0109 & 0.0341 & 0.0135\\
Fixed-depth loss & 0.0642 & 0.0291 & 0.0866 & 0.0347 & 0.0292 & 0.0134 & 0.0435 & 0.0170\\
\midrule
\multicolumn{9}{l}{\emph{RQ-KMeans width and depth:}}\\
\midrule
$J{=}255$, $Q{=}3$ & 0.0648 & 0.0443 & 0.0894 & 0.0505 & 0.0297 & 0.0160 & 0.0435 & 0.0194\\
$J{=}128$, $Q{=}2$ & 0.0623 & 0.0282 & 0.0850 & 0.0339 & 0.0280 & 0.0130 & 0.0421 & 0.0165\\
$J{=}128$, $Q{=}4$ & 0.0621 & 0.0282 & 0.0844 & 0.0337 & 0.0285 & 0.0132 & 0.0426 & 0.0167\\
\bottomrule
\end{tabular}
\end{table}

\subsection{Sensitivity to SID width and depth}

Table~\ref{tab:variants} varies the number of centroids per codebook $J$ at fixed semantic depth $Q=3$, and the number of residual codebooks $Q$ at fixed width $J=128$.

\subsubsection{SID width.}
Increasing $J$ from 128 to 255 leaves NDCG unchanged at both cutoffs on both datasets to the reported precision. Industrial Recall@20 moves from 0.0893 to 0.0894. On Tools, Recall@10 falls from 0.0308 to 0.0297 and Recall@20 from 0.0453 to 0.0435. The additional centroids yield no consistent retrieval gain across the two catalogs.

\subsubsection{SID depth.}
At $J=128$, three codebooks yield the highest Recall and NDCG at both cutoffs on both datasets. For $Q=2$, $3$, and $4$, NDCG@20 is 0.0339, 0.0505, and 0.0337 on Industrial and 0.0165, 0.0194, and 0.0167 on Tools. Each extra codebook gives residual quantization another stage of refinement and adds a semantic step to the path scored by beam search.

Across the tested settings, we discover that Jev-GR framework is more sensitive to depth than to width.

\subsection{Time Cost Analysis}
\label{sec:time-cost}
Table~\ref{tab:time-cost} compares training time per logged epoch and offline test time per user. We divide final test wall time by 9,331 users on Industrial and 16,638 on Tools. Jev-GR uses $J=128$, $Q=3$, and test beam width 20. Its test time includes shard checks and output writing. All training, validation and testing are conducted on 
4 Intel(R) Xeon(R) Silver 4214 CPU @ 2.20GHz (12 cores) CPUs with one Nvidia Geforce RTX 4090 GPU.
\begin{table}[!htb]
\caption{Wall-clock training and inference cost. Epochs are averaged over the recorded runs. Estimated from second-resolution timestamps. Generator training time is estimated after subtracting validation and includes calibration and saving.}
\label{tab:time-cost}
\centering\small\setlength{\tabcolsep}{3pt}
\begin{tabular}{@{}lcccccc@{}}
\toprule
Dataset & Measure & SASRec & BERT4Rec & TIGER & LETTER--TIGER & Jev-GR\\
\midrule
Industrial & $S$/epoch & 1.4098 & 1.5400 & 3.0143 & 7.9667 & 549.6020\\
           & $mS$/user & 0.1072 & 0.1072 & 2.4649 & 3.4294 & 980.0337\\
Tools      & $S$/epoch & 2.7677 & 2.8250 & 6.0000 & 13.9111 & 1064.7320\\
           & $mS$/user & 0.0601 & 0.1202 & 4.2673 & 5.1088 & 968.9740\\
\bottomrule
\end{tabular}
\end{table}

Jev-GR trained for 64.1 minutes over seven epochs on Industrial and 106.5 minutes over six on Tools. Validation accounted for the largest share of each job (67.5\% and 65.2\%, respectively). Final testing took 2\,h 32\,min and 4\,h 29\,min, compared with 23\,s and 71\,s for TIGER. The longer runs retrieved 134 and 143 additional top-20 targets, respectively (Table~\ref{tab:industrial}).

\section{Conclusion}
At a branching prefix, predicting the next SID symbol means choosing which catalog subtree to pursue for a particular user. Jev-GR ranks the legal continuations from a single state encoding the user's history and the current prefix. Its training objective averages the loss over the branching positions of each target path. In full-catalog evaluation on both Amazon datasets, Jev-GR improves Recall@20 and NDCG@20 over the evaluated baselines. At a width of 128 centroids per codebook, three residual codebooks perform best among the tested depths. Freezing the OpenJev-initialized adapter gives the largest decline among the training ablations. Dividing the loss by the full SID length reduces NDCG more than Recall in relative terms, consistent with fork-based normalization improving the ranking of retrieved targets.

\section{Future works}
Jev-GR learns from local choices along the observed item’s SID, while retrieval depends on the cumulative probability of the complete path. In future work, we will investigate Group Sequence Policy Optimization (GSPO) as a post-training step to better align these choices with item-level ranking. For each user history, we will sample catalog-valid SIDs and rank them alongside the observed item’s SID. Exact matches and rank-dependent penalties will provide a group-relative reward for updating the probabilities of sampled paths. We will compare the resulting model with the supervised Jev-GR checkpoint at the same beam width to determine whether this additional training improves Recall and NDCG.

\bibliographystyle{splncs04}
\bibliography{openjev-sid}

\begin{thebibliography}{10}
\providecommand{\url}[1]{\texttt{#1}}
\providecommand{\urlprefix}{URL }
\providecommand{\doi}[1]{https://doi.org/#1}

\bibitem{almeida2026jev}
Almeida, D.: Introducing {System One} models \& {Jev}. TypeSafe AI, September 15 (2026), \url{https://typesafe.ai/blog/introducing-system-one-models-and-jev}

\bibitem{jevbench}
{Benchmark Heaven}: {JevBench}: Benchmark for {AI} decision models. Benchmark and methodology (2026), \url{https://github.com/fstandhartinger/jevbench}

\bibitem{hou2024rank}
Hou, Y., Zhang, J., Lin, Z., Lu, H., Xie, R., McAuley, J., Zhao, W.X.: Large language models are zero-shot rankers for recommender systems. In: Advances in Information Retrieval (ECIR) (2024). \doi{10.1007/978-3-031-56060-6_24}

\bibitem{hu2022lora}
Hu, E.J., Shen, Y., Wallis, P., Allen-Zhu, Z., Li, Y., Wang, S., Wang, L., Chen, W.: {LoRA}: Low-rank adaptation of large language models. In: International Conference on Learning Representations (2022), \url{https://openreview.net/forum?id=nZeVKeeFYf9}

\bibitem{ju2025handbook}
Ju, C.M., Collins, L., Neves, L., et~al.: Generative recommendation with {Semantic IDs}: A practitioner's handbook. In: Proceedings of the ACM International Conference on Information and Knowledge Management (2025). \doi{10.1145/3746252.3761612}

\bibitem{kang2018sasrec}
Kang, W.C., McAuley, J.: Self-attentive sequential recommendation. In: IEEE International Conference on Data Mining. pp. 197--206 (2018). \doi{10.1109/ICDM.2018.00035}

\bibitem{kong2025minionerec}
Kong, X., Sheng, L., Tan, J., Chen, Y., Wu, J., Zhang, A., Wang, X., He, X.: {MiniOneRec}: An open-source framework for scaling generative recommendation. arXiv:2510.24431 (2025), \url{https://arxiv.org/abs/2510.24431}

\bibitem{lee2025gram}
Lee, S., Choi, M., Choi, E., Kim, H., Lee, J.: {GRAM}: Generative recommendation via semantic-aware multi-granular late fusion. In: Proceedings of the Annual Meeting of the Association for Computational Linguistics (2025), \url{https://aclanthology.org/2025.acl-long.1596/}

\bibitem{li2026jeva}
Li, Q., Wang, X.: Jeva: System one models from first principles. arXiv preprint arXiv:2610.18773  (2026)

\bibitem{li2024generativesearch}
Li, Y., Lin, X., Wang, W., Feng, F., Pang, L., Li, W., Nie, L., He, X., Chua, T.S.: A survey of generative search and recommendation in the era of large language models. arXiv:2404.16924 (2024), \url{https://arxiv.org/abs/2404.16924}

\bibitem{liu2026scaling}
Liu, J., Collins, L., Tang, J., et~al.: Understanding generative recommendation with {Semantic IDs} from a model-scaling view. In: Proceedings of the ACM SIGKDD Conference on Knowledge Discovery and Data Mining (2026), \url{https://arxiv.org/abs/2509.25522}

\bibitem{morin2005hierarchical}
Morin, F., Bengio, Y.: Hierarchical probabilistic neural network language model. In: Proceedings of the Tenth International Workshop on Artificial Intelligence and Statistics (2005), \url{https://proceedings.mlr.press/r5/morin05a.html}

\bibitem{ni2019amazon}
Ni, J., Li, J., McAuley, J.: Justifying recommendations using distantly-labeled reviews and fine-grained aspects. In: Proceedings of EMNLP-IJCNLP. pp. 188--197 (2019). \doi{10.18653/v1/D19-1018}

\bibitem{openjevcard}
{Open-Jev contributors}: {Open-Jev-2B} model card. Hugging Face model release (2026), \url{https://huggingface.co/ZefanCai/Open-Jev-2B}

\bibitem{openjev}
{Open-Jev contributors}: {Open-Jev}: Open-source candidate scoring for typed decisions. \url{https://github.com/Zefan-Cai/Open-Jev} (2026)

\bibitem{qwen}
{Qwen Team}: {Qwen3.5-2B} model card. \url{https://huggingface.co/Qwen/Qwen3.5-2B} (2026)

\bibitem{rajput2023tiger}
Rajput, S., Mehta, N., Singh, A., Keshavan, R.H., Vu, T., Heldt, L., Hong, L., Tay, Y., Tran, V.Q., Samost, J., Kula, M., Chi, E.H., Sathiamoorthy, M.: Recommender systems with generative retrieval. In: Advances in Neural Information Processing Systems (2023), \url{https://arxiv.org/abs/2305.05065}

\bibitem{sun2019bert4rec}
Sun, F., Liu, J., Wu, J., Pei, C., Lin, X., Ou, W., Jiang, P.: {BERT4Rec}: Sequential recommendation with bidirectional encoder representations from transformer. In: Proceedings of the 28th ACM International Conference on Information and Knowledge Management. pp. 1441--1450 (2019). \doi{10.1145/3357384.3357895}

\bibitem{tsaruk2026preference}
Tsaruk, V., Gusak, D., Korst, R., Frolov, E.: Preference-aligned {Semantic IDs} for generative recommendation. In: Proceedings of the ACM International Conference on Information and Knowledge Management (2026). \doi{10.1145/3799682.3840019}

\bibitem{typesafehierarchy}
{TypeSafe AI}: Hierarchical classification. TypeSafe cookbook (2026), \url{https://docs.typesafe.ai/cookbooks/hierarchical_classification}

\bibitem{typesafechoice}
{TypeSafe AI}: {System One}. Technical documentation (2026), \url{https://docs.typesafe.ai/concepts/system-one}

\bibitem{wang2024letter}
Wang, W., Bao, H., Lin, X., Zhang, J., Li, Y., Feng, F., Ng, S.K., Chua, T.S.: Learnable item tokenization for generative recommendation. In: Proceedings of the 33rd ACM International Conference on Information and Knowledge Management (2024). \doi{10.1145/3627673.3679569}

\bibitem{wang2022nci}
Wang, Y., Hou, Y., Wang, H., Miao, Z., Wu, S., Sun, H., Chen, Q., Xia, Y., Chi, C., Zhao, G., Liu, Z., Xie, X., Sun, H.A., Deng, W., Zhang, Q., Yang, M.: A neural corpus indexer for document retrieval. In: Advances in Neural Information Processing Systems (2022), \url{https://arxiv.org/abs/2206.02743}

\bibitem{wu2024llmrec}
Wu, L., Zheng, Z., Qiu, Z., Wang, H., Gu, H., Shen, T., Qin, C., Zhu, C., Zhu, H., Liu, Q., Xiong, H., Chen, E.: A survey on large language models for recommendation. World Wide Web  \textbf{27}(5) (2024). \doi{10.1007/s11280-024-01291-2}

\bibitem{yu2026daso}
Yu, X., Li, S., Aghaei, S., Zhu, Z., Bai, J., Huang, G., Peng, B., Liu, Y., Xue, L.: Difficulty-aware {Semantic-ID} optimization for generative recommendation. arXiv:2608.20611 (2026), \url{https://arxiv.org/abs/2608.20611}

\bibitem{zheng2024lcrec}
Zheng, B., Hou, Y., Lu, H., et~al.: Adapting large language models by integrating collaborative semantics for recommendation. In: IEEE International Conference on Data Engineering (2024), \url{https://arxiv.org/abs/2311.09049}

\bibitem{zhu2018tdm}
Zhu, H., Li, X., Zhang, P., Li, G., He, J., Li, H., Gai, K.: Learning tree-based deep model for recommender systems. In: Proceedings of the 24th ACM SIGKDD International Conference on Knowledge Discovery and Data Mining (2018). \doi{10.1145/3219819.3219826}

\end{thebibliography}
\end{document}